\documentclass[pra,twocolumn,superscriptaddress,preprintnumbers]{revtex4-1}

\usepackage{epsfig}
\usepackage{amsmath}
\usepackage{amssymb}
\usepackage{amsfonts}
\usepackage{mathptmx}
\usepackage{dcolumn}
\usepackage{eucal}
\usepackage{bm,bbm}
\usepackage{color}
\usepackage[colorlinks,linkcolor=blue,citecolor=blue]{hyperref}
\usepackage{bbold}
\usepackage{dsfont}
\usepackage{url}

\newcommand{\wcl}{W_\textrm{C}}
\newcommand{\vq}{V_\textrm{Q}}
\newcommand{\ket}[1]{|#1\rangle}
\newcommand{\bra}[1]{\langle#1|}
\newcommand{\braket}[1]{\langle#1\rangle}
\newcommand{\dm}[1]{\ket{#1}\bra{#1}}
\newcommand{\bfx}{\mathbf{x}}
\newcommand{\bfp}{\mathbf{p}}

\newcommand{\D}{\hat{D}}
\def\tr{\mathrm{Tr}}
\def\d{\mathrm{d}}
\def\uj{{}^{(j)}}
\def\un{{}^{(n)}}
\def\um{{}^{(m)}}
\def\upsi{{}^{(\psi)}}
\def\upsit{{}^{(\psi_{\rm tot})}}

\newcommand{\U}{\hat{U}_{\tau}}
\newcommand{\atxt}{|_{(\bfx,t)}}
\def\lpsij{{}_{\{\psi\uj\}}}

\begin{document}


\title{Quantum Work in the Bohmian framework}

\author{R. Sampaio}
\email[]{rui.ferreirasampaio@aalto.fi}

\author{S. Suomela}
\affiliation{COMP Center of Excellence, Department of Applied Physics, Aalto University, P.O. Box 11000, FI-00076 Aalto, Finland.}

\author{T. Ala-Nissila}
\affiliation{COMP Center of Excellence, Department of Applied Physics, Aalto University, P.O. Box 11000, FI-00076 Aalto, Finland.}
\affiliation{Departments of Mathematical Sciences and Physics, Loughborough University, Loughborough,  
Leicestershire LE11 3TU, United Kingdom.}

\author{J. Anders}

\author{T. G. Philbin}
\email[]{t.g.philbin@exeter.ac.uk}
\affiliation{CEMPS, Physics and Astronomy, University of Exeter, Exeter, EX4 4QL, United Kingdom.}

\begin{abstract}

At non-zero temperature classical systems exhibit statistical fluctuations of thermodynamic quantities arising from the variation of the system's initial conditions and its interaction with the environment. The fluctuating work, for example, is characterised by the ensemble of system trajectories in phase space and, by including the probabilities for various trajectories to occur, a work distribution can be constructed. However, without phase space trajectories, the task of constructing a work probability distribution in the quantum regime has proven elusive. Here we use quantum trajectories in phase space and define fluctuating work as power integrated along the trajectories, in complete analogy to classical statistical physics. The resulting work probability distribution is valid for any quantum evolution, including cases with coherences in the energy basis. We demonstrate the quantum work probability distribution and its properties with an exactly solvable example of a driven quantum harmonic oscillator. An important feature of the work distribution is its dependence on the initial statistical mixture of pure states, which is reflected in higher moments of the work. The proposed approach introduces a fundamentally different perspective on quantum thermodynamics, allowing full thermodynamic characterisation of the dynamics of quantum systems, including the measurement process.

\end{abstract}

\date{\today} 

\maketitle


\section{Introduction}

Landau said ``All the concepts and quantities of thermodynamics follow most naturally, simply and rigorously from the concepts of statistical physics'' \cite{LandauLifschitz}. While the second law of thermodynamics puts limits on the work drawn from a system, this work is most naturally viewed as the average over a statistical work distribution \cite{Bochkov1977,Jarzynski1997}.
Classically, the work distribution is constructed from the ensemble of system trajectories in phase space. These trajectories specify the energy of the system at all times, thus allowing the work done on the system to be computed. In the quantum case the concept of fluctuating work has proven elusive \cite{Allahverdyan2005, Vinjanampathy16} because, in the conventional view, trajectories in quantum mechanics are considered to be impossible~\cite{Bell1982}. Trajectories in phase space can, however, be constructed in an alternative formulation of quantum mechanics \cite{deBroglie1925,Bohm:1952aa,Bohm:1952ab,Bell1982,holland1995quantum,Buchanan2011} which makes predictions consistent with experimental results. Here we utilise this approach to define work and its distribution function for quantum systems.

Numerous definitions of quantum work have been proposed~\cite{Lindblad1983, Yukawa2000, Tasaki2000, Kurchan2000, Pekola2013, Alonso2016, Deffner2015, Aberg2013a, Kammerlander2016, Hayashi2016, Chernyak2004, Venkatesh2015, Allahverdyan2014a, Miller2016, Solinas2016, Talkner2016, SolinasMiller2017,Campisi2013,Deffner2016,Esposito2006,deffner2013quantum}, the most widely used being the two measurement protocol (TMP) definition for closed quantum systems~\cite{Tasaki2000, Kurchan2000, Talkner2007}. The TMP leads to a quantum version of the Jarzynski equality~\cite{Jarzynski1997, Bochkov1977, Tasaki2000, Mukamel2003} and the Tasaki-Crooks relation~\cite{Tasaki2000, Talkner2007a, Cohen2012}, the classical correspondence has been elucidated~\cite{Jarzynski2015} and experimental implementation is relatively straightforward~\cite{An2014,Pekola2013,Huber2008,Batalhao2015,Campisi2013a,PhysRevX.4.031029,Dorner2013,Cerisola2017}. The TMP has been extended to open systems~\cite{Campisi2009,Hekking:2013aa,Suomela:2014aa,deffner_quantum_2011,Salmilehto2014}, to continuously measured processes~\cite{Campisi2010, Watanabe2014a} and to relativistic systems~\cite{deffner_quantum_2015}.
Other approaches include using a power or work operator~\cite{Lindblad1983, Yukawa2000, Chernyak2004,Allahverdyan2005, Engel2007, Solinas2013}, measuring the system by coupling it weakly to a detector~\cite{Alonso2016}, the full counting statistics (FCS) approach~\cite{Solinas2016,Talkner2016,SolinasMiller2017}, and work definitions based on entropic principles~\cite{Aberg2013a, Lostaglio15, Kammerlander2016, Hayashi2016, PerarnauRiera16, Alhambra2016, richens_work_2016}. Work definitions based on trajectories in Hilbert space include the quantum jump approach~\cite{Hekking:2013aa,Horowitz2012,Elouard17} and the consistent histories framework~\cite{Miller2016}. 
None of these approaches yields a positive work distribution based on trajectories in phase space. 

In this paper we provide a positive work distribution for arbitrary statistical mixtures of wave functions in a manner fully analogous to the classical work definition. Contrasting with all previous studies of work in the quantum regime, this work distribution is established using quantum Hamilton-Jacobi theory which we briefly recall in Sec.~\ref{sec:QHJ}. In Sec.~\ref{sec:work} we show how the fluctuating work is defined for initial pure states, i.e. wave functions, and derive the work distribution for closed processes that begin in an initial statistical mixtures of wave functions. We show that the average work of this distribution always coincides with the change in the expectation value of the internal energy, i.e. energy conservation is ensured. The relation to the TMP work distribution and the use of the Hamilton-Jacobi approach for finite-dimensional systems are also discussed. In Sec.~\ref{sec:example} the Hamilton-Jacobi work distribution is illustrated with an exactly solvable example, a driven quantum harmonic oscillator. We conclude in Sec.~\ref{sec:conclusion} that the work distribution based on Hamilton-Jacobi theory provides a natural characterisation of work and its fluctuations in coherent quantum systems while recovering the classical definition in the high-temperature limit. We highlight a number of open questions that pertain to the Hamilton-Jacobi approach to work in the quantum regime, such as initial entangled states and open dynamics.
An important observation is that, in general, the work distribution associated with closed quantum dynamics will depend on the preparation of the initial mixture.

\section{Quantum Hamilton-Jacobi theory.} \label{sec:QHJ} 
Quantum mechanics can be formulated as a theory of trajectories in phase space as shown by Bohm and others \cite{deBroglie1925, Bohm:1952aa, Bohm:1952ab, DurrQuantum12,holland1995quantum,teufel2009bohmian}.
Writing the wave function for a single particle in polar form, $\psi(\bfx, t) = R(\bfx, t) \, \exp(i S(\bfx, t)/\hbar)$, where $R \ge 0$ is the amplitude and $S \in \mathbbm{R}$ is the phase of the wave function, and $\bfx$ is the particle position, the Schr\"odinger equation reduces to two coupled equations for $R$ and $S$. The imaginary part gives the continuity equation for the probability density $R^2$. The real part has the form of the Hamilton-Jacobi equation 
\begin{equation}\label{eq:EBohm1}
	\frac{\partial S\upsi(\bfx,t)}{\partial t} + E\upsi (\bfx, t) = 0,
\end{equation}
where 
\begin{equation}\label{eq:EBohm2}
	E\upsi(\bfx, t)  = \frac{\bfp\upsi(\bfx,t)^2}{2m} + V(\bfx, \bfp\upsi(\bfx,t), t) + \vq\upsi(\bfx,t).
\end{equation}
Here $m$ is the mass of the particle, $V(\bfx, \bfp\upsi(\bfx,t), t)$ is the external potential and $ \bfp\upsi(\bfx,t) \equiv \nabla S\upsi\atxt $. The above is exactly the classical equation for the same problem but with an additional term, the quantum potential $\vq\upsi \equiv -\hbar^2 (\nabla^2 R\upsi)/2m R\upsi$.  
According to Hamilton-Jacobi theory, $E\upsi(\bfx, t) $ and $ \bfp\upsi(\bfx,t) $ are the energy and canonical momentum at point $ \bfx $ and time $ t $, respectively.
To obtain the trajectories we integrate Hamilton's equation of motion $\dot{\bfx}_t = \partial_{\bfp}\mathcal{H}\upsi|_{(\bfx_t,\bfp\upsi(\bfx_t,t),t)} $ where $\bfx_t \equiv \bfx (t)$ and $ \mathcal{H}\upsi(\bfx,\bfp,t) \equiv \bfp^2/2m + V(\bfx,\bfp,t) + V\upsi_\mathrm{Q}(\bfx,t)$. 

\begin{figure}[t]
    \includegraphics[width=0.85\linewidth]{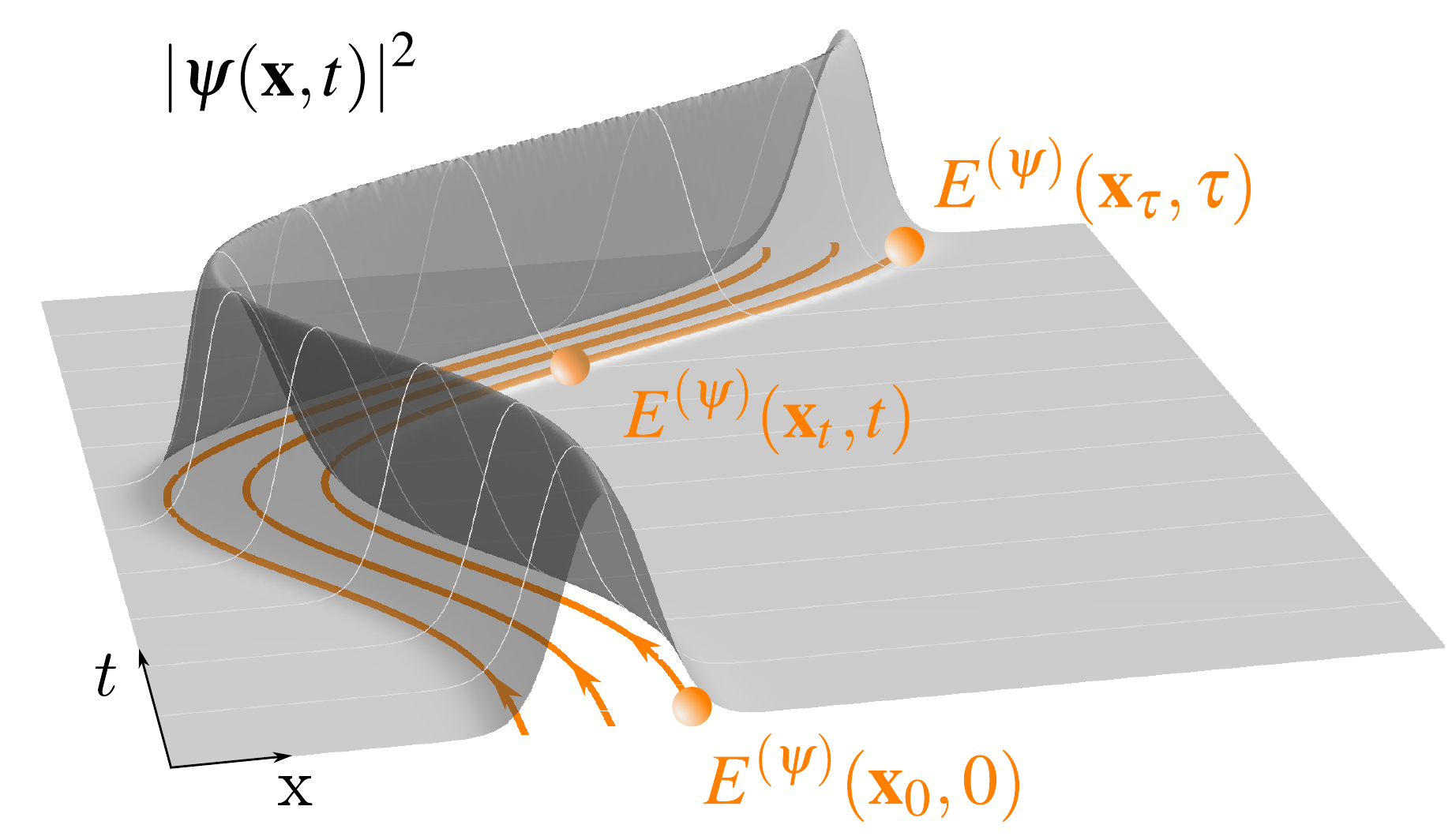}
    \caption{    \label{fig1}
Bohmian trajectories for a driven 1D harmonic oscillator, see main text for details. The orange lines show three possible trajectories of the particle all belonging to the ensemble for the wave function $\psi(\bfx, t)$ that starts in the lowest energy eigenstate. The trajectories are specified by the particle's initial position, $\bfx_0$. Also shown is $|\psi(\bfx, t)|^2$ (grey surface); this represents the probability density for finding the particle on each trajectory of the ensemble. The energy of the particle at any point of the trajectory is $E\upsi(\bfx_t, t)$. 
}
\end{figure}

For each wave function $\psi(\bfx, t)$, this gives an ensemble of trajectories, namely one trajectory for each initial position $\bfx_0$ of the particle, see Fig.~\ref{fig1} for an example. Experimentally these particle trajectories have been reconstructed using weak measurements~\cite{Kocsis2011, Mahler2016, Xiao2017}. The probability for being on a trajectory with initial position $\bfx_0$ inside an infinitesimal volume $\d\bfx_0$ is given by  $R^2(\bfx_0, 0)\d\bfx_0 = |\psi (\bfx_0, 0)|^2\d\bfx_0$, {\it i.e.} the Born rule \cite{Philbin15,holland1995quantum}. The continuity equation ensures that  the Born rule distribution of the trajectories is preserved in time as $|\psi (\bfx_t, t)|^2\d \bfx_t = |\psi (\bfx_0, 0)|^2\d\bfx_0$, where $ \d\mathbf{x}_t $ is the time evolved initial infinitesimal volume $ \d\mathbf{x}_0 $. Note that whenever the quantum potential can be neglected from the total energy, the classical Hamiltonian and the corresponding equations of motion are recovered.

\section{Work in quantum Hamilton-Jacobi theory.} \label{sec:work} 

\subsection{Fluctuating work definition.} 

For each wave function $\psi(\bfx, t)$ one can now define work as power integrated along any trajectory $\bfx_t$ just as in classical mechanics, i.e. the work done on the system between time $0$ and $\tau$ is 
\footnote{An alternative work definition would follow if the quantum potential is dropped in Eq.~\eqref{eq:EBohm2}, the reduced energy is denoted by $E_{\rm C}\upsi(\bfx,t) = E\upsi(\bfx,t) - V_\mathrm{Q}\upsi(\bfx, t)$ and the work is identified as $\wcl\upsi[\bfx_t] \equiv \int_0^\tau \d t\,\partial \mathcal{H}_{\rm C}\upsi/\partial t|_{\bfx_t}$, with $ \mathcal{H}_{\rm C} \equiv \mathcal{H} - V\upsi_\mathrm{Q}$. However, this work $\wcl\upsi[\bfx_t]$ is not the same as the classical expression because the trajectory will still depend on the quantum potential. Furthermore, since $\mathcal{H}_{\rm C}\upsi$ does not generate the particle dynamics even in the closed case, the work $\wcl\upsi[\bfx_t]$ is not equal to the change in $E_{\rm C}\upsi$ along a quantum trajectory.}
\begin{equation}\label{eq:work1} 
	W\upsi[\bfx_t]  \equiv \int_0^{\tau} \d t~\frac{\partial \mathcal{H}\upsi}{\partial t}\Big|_{\bfx_t}.
\end{equation}
Under unitary evolution Eq.~\eqref{eq:work1} correctly reduces to the energy difference between the end points of the trajectory $W\upsi[\bfx_t] = E\upsi (\bfx_{\tau},\tau) - E\upsi(\bfx_0,0)$. 
For open dynamics, a full integration over the trajectory would have to be performed, using the total wave function, $\psi_{\rm tot}$, 
of the system and degrees of freedom to which it couples. If a system's open dynamics arises from it coupling to a large reservoir that has a well defined temperature, then the difference between the energy change, $E\upsit (\bfx_{\tau},\tau) - E\upsit(\bfx_0,0)$, and the fluctuating work, $W\upsit[\bfx_t]$, can be identified with ``fluctuating heat'' absorbed by the system from the reservoir.

Given definition \eqref{eq:work1}, the probability distribution for quantum work when the system starts in an initial pure state $\psi(\bfx,0)$ is simply
\begin{equation}  \label{PtauW}
    P(W;\psi) = \int \d\bfx_0 \, |\psi(\bfx_0,0)|^2 \, \delta(W - W\upsi[\bfx_t]).
\end{equation}
This is the work distribution assuming that the system is initially described by a single wave function and undergoes either closed or open dynamics. We now discuss the case where the system starts in a statistical mixture of pure states, $\{\psi\uj\}$, with the index $j$ taken to be discrete for simplicity. We refer to statistical mixtures as those that have been prepared in a statistical manner - i.e. with probability $p_j$ the pure state $\psi\uj$ was prepared. The probability distribution of work for this statistical mixture is then given by
\begin{equation}\label{eq:p(w)} 
    P_{\{\psi\uj\}}(W) = \sum_j p_j  \, P(W;\psi\uj).
\end{equation}
This is a positive probability distribution, i.e. $P_{\{\psi\uj\}}(W) \geq 0$ for all $W$ and $\int \d W \, P_{\{\psi\uj\}}(W)  = 1$. It is important to note that $P_{\{\psi\uj\}}(W)$ depends explicitly on the mixture of wave functions  $\{\psi\uj\}$. 

Our definition of $P_{\{\psi\uj\}}(W)$ above means that two different statistical mixtures will produce, in  general, two different probability distributions of work even when they correspond to the same density operator, $\hat\rho$. We give an explicit example of the mixture dependence of the work probability distribution below. If the system is initially entangled with other degrees of freedom the reduced state of the system will be mixed but it does not correspond to a proper physical mixture of pure states. The work distribution is then not given by \eqref{eq:p(w)} as there is no unique set of mixing probabilities. In this case, the full wave function $\psi_{\rm tot}$ describing the system and correlated degrees of freedom is required and the work distribution is given by $P(W;\psi_{\rm tot})$, see Eq.~\eqref{PtauW}. For a discussion of the distinction between statistical mixtures and density operators see, for example,~\cite{Allahverdyan2005,Kok2000,breuer2002theory,Goldstein2006}. In this paper, we focus on the case where the initial state of the system is a proper statistical mixture of pure states. 

\subsection{Average work.} 

For a statistical mixture  $\hat\rho$ of pure states, $\{\psi\uj\}$ with probabilities $\{p_j\}$, the average work under unitary evolution can be obtained from the distribution \eqref{eq:p(w)} as $\braket{W} \equiv \int \d W P_{\{\psi\uj\}}(W) \, W$, explicitly,
\begin{equation}\label{avg-work}
	\begin{aligned}
	& \braket{W}\lpsij \\
	& = \sum_j p_j \int \d \bfx_0 \, |\psi\uj(\bfx_0,0)|^2 \, (E\uj(\bfx_\tau,\tau) - E\uj(\bfx_0,0)) \\
	&= \sum_j p_j \int \d \bfx_0 \, |\psi\uj(\bfx_0,0)|^2 \, E\uj(\bfx_\tau,\tau) - \braket{E(\bfx,0)}\lpsij \\
	& = \sum_j p_j \int \d \bfx_{\tau} \, |\psi\uj(\bfx_{\tau},\tau)|^2 \, E\uj(\bfx_{\tau},\tau) - \braket{E(\bfx,0)}\lpsij \\
	& = \braket{E(\bfx,\tau)}\lpsij - \braket{E(\bfx,0)}\lpsij \\
	& = \mathrm{Tr}[\hat{H}(\tau)\hat\rho(\tau)] - \mathrm{Tr}[\hat{H}(0)\hat\rho(0)].
	\end{aligned}
\end{equation}
Here we have used the short-hand notation $\braket{E(\bfx,t)}\lpsij \equiv \sum_j p_j \int\mathrm{d}\bfx_t|\psi\uj(\bfx_t,t)|^2 \, E\uj(\bfx_t,t)$ in lines 2-4, and the equivariance property $\d \bfx_0 \, |\psi\uj(\bfx_0,0)|^2 = \d \bfx_{\tau} \, |\psi\uj(\bfx_{\tau},\tau)|^2$ in line 3.  We have used the identity $E\uj(\bfx,t) \equiv \Re\{(H\psi\uj)(\bfx,t)/\psi\uj(\bfx,t)\}$  in line 5~\cite{holland1995quantum} and thus for $\hat\rho (t) = \sum_j p_j \dm{\psi\uj(t)}$ one obtains $\braket{E(\bfx,t)}\lpsij = \sum_j p_j \int\d\bfx |\psi\uj(\bfx,t)|^2 E\uj(\bfx,t) = \tr [\hat\rho(t) \hat{H}(t)]$ at time $t$. 

Thus, for a closed system one always finds that $\braket{W}$ is equal to the change in  the expectation value of the Hamiltonian operator of the system, $\hat{H}(t)$, i.e. $\braket{W} = \tr [\hat{H}(\tau) \, \U \, \hat\rho \, \U^{\dag}] - \tr [\hat{H}(0) \, \hat\rho]$ where $\hat\rho$ is the density operator for the initial statistical  mixture and $\U$ is the evolution operator for the time-interval $\tau$. 
Note that because of the linearity of the averages the choice of how the initial density operator $\hat\rho$ was mixed plays no role, i.e. for any set of pure states $\{\psi\uj\}$ that corresponds to the density matrix $\hat\rho$ the same average work is obtained. We highlight that, in general, this will \emph{not} be the case for higher moments of work, such as the average exponentiated work (see below).

\subsection{Measurements and the TMP.}
A measurement implies a coupling of the system to a measuring device leading to open dynamics of the system during the time interval of the measurement.
In the Bohmian formulation measurements can be explicitly included as part of the continuous and deterministic dynamics of the system and the measuring device~\cite{holland1995quantum,DurrQuantum12}. The system still follows a trajectory in phase space and so the work can again be computed as integrated power along this trajectory.  A general discussion of the thermodynamics of measurements lies outside the scope of this paper, but it is pertinent to understand how the work definition \eqref{eq:work1} relates to the TMP definition. The protocol in TMP includes two energy measurements of the system~\cite{Talkner2007}. We now describe the TMP in the Hamilton-Jacobi picture and then discuss circumstances in which the TMP work distribution will coincide with that 
of Eq. \eqref{eq:work1}.

The first energy measurement on an initial density operator $\hat\rho$ dynamically selects one of the eigenstates $\phi^{(n)}(\bfx, 0)$ of the Hamiltonian $\hat{H}(0)$ with eigenvalue $E\un_0$. This happens in a dynamical process \cite{holland1995quantum, DurrQuantum12} in which the measuring device moves to give an outcome $E\un_0$ which is associated with a system trajectory belonging to the ensemble of trajectories of the wave function $\phi^{(n)}(\bfx, 0)$.
The probability for obtaining outcome $E\un_0$ in this measurement is  $q_n = \bra{\phi^{(n)}} \, \hat\rho \, \ket{\phi^{(n)}}$. The negative time derivative of the phase of the wave function $\phi^{(n)}(\bfx, t)$ at time $t=0$ is $E\un_0$. Hence the particle energy  $E^{(\phi^{(n)})}(\bfx, 0)$ in Eq.~\eqref{eq:EBohm1} after the first measurement is just the energy eigenvalue $E\un_0$, for any position of the particle. The system now evolves according to the Hamiltonian $\hat{H}(t)$ such that at time $\tau$ the wave function of the system is $\U \, \phi^{(n)}(\bfx, 0)$. 

At this time another measurement of the energy is performed. As for the first measurement, this second measurement  dynamically evolves the system to an eigenstate $\chi^{(m)} (\bfx, \tau)$ of the Hamiltonian $\hat{H}(\tau)$ with eigenvalue $E\um_{\tau}$.
The probability of obtaining this outcome $E\um_{\tau}$, given the first outcome $E\un_{0}$, is the conditional probability $p_{m|n}=| \bra{\chi^{(m)}} \U \, \ket{\phi^{(n)}}  |^2$. 
The distribution $P_{\hat\rho}(\Delta E)$ for the change in energy outcomes is given by   $P_{\hat\rho}(\Delta E) = \sum_n q_n  \, P(\Delta E;\phi\un)$ with $P(\Delta E;\phi\un) = \sum_m p_{m|n} \, \delta(\Delta E - (E\um_\tau - E\un_0))$.  Since in TMP the energy change is identified as work this is the TMP work distribution.

If we now consider the same protocol, but with work defined as in Eq.\ \eqref{eq:work1}, then the integrated power must be computed along the entire system trajectory, including the time intervals where the two measurements take place.
Any work done on the system \emph{during} the two measurements will thus be included, in contrast to TMP. Thus the work distribution based on \eqref{eq:work1} will in general differ from the TMP work distribution $P_{\hat\rho}(\Delta E)$. This difference arises due to the fact that during the measurements the dynamics of the system is actually open, as has been pointed out previously by Kammerlander {\it et al.}, Solinas {\it et al.}, and Elouard {\it et al.} \cite{Kammerlander2016,Solinas15,Elouard17}.
There are circumstances, however, where the TMP work definition will agree with the trajectory definition \eqref{eq:work1}. For example, if the system is in a mixture of energy eigenstates before each energy measurement, and if in addition the measurements make a negligible contribution to the work compared to the rest of the protocol, then the work \eqref{eq:work1} will be equal to the difference in the energy outcomes, in agreement with the TMP  work definition.

\subsection{Finite-dimensional systems.}
While we have here considered particles with continuous variables, discrete systems, such as qubits, can also be treated in the Hamilton-Jacobi approach~\cite{holland1995quantum,DurrQuantum12,Roser12}. For example, if a qubit is physically realised by a spin-1/2 particle then one sets up a spinor wave function and then extracts one overall phase from it, which determines the ensemble of particle trajectories. If a qubit is taken as the lowest two energy levels of a spin-less particle, then the formulation described here, using a scalar wave function, is sufficient. 
As an example, suppose the qubit is realised as the two lowest levels of a particle in a one-dimensional infinite-well potential of width $L$ \cite{Roser12}. The wave function of any given pure state of the qubit is given by $\psi(x,t) = c_0(t)\phi_0(x) + c_1(t)\phi_1(x)$, where $c_0(t)$ and $c_1(t)$ are time-dependent complex valued functions and $\phi_0(x)=\sqrt{2/L}\sin(\pi x/L)$ is the ground state and $\phi_1(x)=\sqrt{2/L}\sin(2\pi x/L)$ is the exited state. The work associated with any operation can then be calculated from Eq. (\ref{eq:work1}). For quantitative details of a spin-1/2 particle in the Hamilton-Jacobi formulation, we refer the reader to the textbooks~\cite{holland1995quantum,DurrQuantum12}.

\section{Example: the driven quantum oscillator.} \label{sec:example} 
To illustrate the work distributions we here consider a quantum harmonic oscillator of mass $m$ and frequency $\omega$ with external driving, described by the Hamiltonian
\begin{equation}\label{aHho}
	\hat{H}(\hat{x}, \hat{p}, t) 
	= \frac{\hat{p}^2}{2m} + \frac{1}{2} m\omega^2 \hat{x}^2- \hat{x}f_1(t) - \hat{p}f_2(t)
\end{equation}
where the last two terms in Eq.~(\ref{aHho}) represent the external driving, given by,
\begin{equation}
	f_1(t)=-A\sin(\omega t), \qquad f_2(t)=-\frac{A}{m\omega}\cos(\omega t),   \label{af2}
\end{equation}
where $A$ is a real constant. 
The external driving is chosen such that an exact solution for the time-evolution operator can be found~\cite{Merz}, which is given by
\begin{equation}
\hat{T}(t,0)=\exp\left\{\frac{At}{\sqrt{2\hbar m\omega}}\left[- \hat{a} e^{i\omega t}+ \hat{a}^\dagger e^{-i\omega t}\right]\right\} \, e^{-i \hat{H_0} t/\hbar} ,  \label{aT2}
\end{equation}
where $\hat{H_0}=\hbar\omega\left(\hat{a}^\dagger \hat{a}+\frac{1}{2}\right)$ is the free-oscillator Hamiltonian. The particle trajectories $x(t)$ are given by the solution of
\begin{equation} \label{axdot}
\dot{x}(t)=\frac{1}{m}\frac{\partial}{\partial x} S(x,t)-f_2(t),
\end{equation}
where we must solve for the time evolution of the wave function to obtain the phase $S(x,t)$. The energy eigenstates are most easily found by rewriting Eq.~(\ref{aHho}) with Eq.~(\ref{af2}) as
\begin{equation}
	\hat{H}(t) = \hbar\omega\left(\hat{b}^\dagger \hat{b}+\frac{1}{2}\right)-\frac{A^2}{2m\omega^2},  \label{Hho}
\end{equation}
with $ \hat{b}=\hat{a}-\alpha e^{-i\omega t} $ and $ \alpha= -iA / \sqrt{2\hbar m\omega^3} $.
In this form it is easy to show that the eigenstates of $\hat{H}(t)$ at fixed $t$ are displaced number states, i.e.\ states obtained by acting with the displacement operator on the energy eigenstates of the \emph{undriven} oscillator (for details and literature on displaced number states, see~\cite{gcstates} for example). In our case the $t=0$ energy eigenstates are displaced number states  $\ket{\tilde{n}_{\alpha}} \equiv \D(\alpha) \ket{n}$, where $\ket{n}$ is the $n$-th energy eigenstate of the undriven oscillator and $\D(\alpha)$ is the displacement operator.  The energy eigenvalues of $\hat{H}(t)$ are $\hbar\omega\left(n+\frac{1}{2}  -|\alpha|^2 \right)$ at all times.

\subsection{Initial thermal mixture of energy eigenstates}

We consider first the case where the system is in an energy eigenstate $\ket{\tilde{n}_{\alpha}}$ at $t=0$. Because of the explicit time dependence in Eq.~(\ref{aHho}), the system does not remain in an energy eigenstate. The time evolution operator (\ref{aT2}) contains the displacement operator, and evolves the initial state $\ket{\tilde{n}_{\alpha}}$ to $e^{iA^2t/(2\hbar m\omega^2)}  \D(\alpha+i\omega t \alpha) \ket{n}$, which is another displaced number state but with $t$-dependent complex amplitude. The evolved state is not an energy eigenstate (for $t>0$). The phase $S(x,t)$ of the wave function for $e^{iA^2t/(2\hbar m\omega^2)}  \D(\alpha+i\omega t \alpha) \ket{n}$ is straightforwardly obtained from the displaced number state results~\cite{gcstates} and takes the form
\begin{align}
	S(x,t)=	& -\hbar\omega\left(n+\frac{1}{2}\right)t   +\frac{A^2 t}{2m\omega^2} \nonumber  \\
			& - \frac{A}{\omega}x\left[\cos(\omega t)+\omega t \sin(\omega t)\right] \nonumber\\
			& + \frac{A^2}{4m\omega^3}\left[2\omega t\cos(2\omega t)+(\omega^2 t^2-1) \sin(2\omega t)\right].  \label{aSen}
\end{align}
The quantum trajectories of the particle are then found from Eq.~(\ref{axdot}) to be
\begin{equation}
	x_n(t)=x_0+\frac{A}{m\omega^2}\left[\omega t \cos(\omega t)-\sin(\omega t)\right],  \label{ahotraj}
\end{equation}
with $x_0$ the initial position. Note that the set of trajectories is the same for each initial energy eigenstate $\ket{\tilde{n}_{\alpha}}$, i.e. Eq.~\eqref{ahotraj} is independent of $n$. The work along a trajectory calculated from Eqs. (\ref{eq:work1}), (\ref{aSen}) and (\ref{ahotraj}) is given by
\begin{equation}   \label{awq2}
	W[x_n(t)]=\frac{A \tau  [A \tau +2 m x_0 \omega  \cos(\omega \tau)]}{2 m}.
\end{equation}
We obtain the work distribution by taking into account the probability of the particle following a trajectory (\ref{ahotraj}) with initial position $x_0$. The probability distribution for the initial position is given by $|\psi_n(x_0)|^2$, where $\psi_n(x)$ is the initial wave function for the $t=0$ energy eigenstate  $\ket{\tilde{n}_{\alpha}}$. The initial position distribution is
\begin{align}
	|\psi_n(x_0)|^2=  &  \frac{1}{2^{n} \, n!} \sqrt{\frac{m\omega}{\pi\hbar}} \, \exp\left\{-\frac{m\omega x_0^2}{\hbar} \right\}  \, \left[ H_n\left(  \sqrt{\frac{m\omega}{\hbar}} \, x_0 \right) \right]^2,  \label{ahox0}
\end{align}
which is the same as that for the energy eigenstates of the undriven oscillator (at $t=0$ the probability density  is not displaced relative to the number state). The probability distribution of work for the initial eigenstate (\ref{ahox0}) is then given by Eq.~\eqref{PtauW}.

We now consider an initial statistical mixture of energy eigenstates at $t=0$ with thermal probabilities,
\begin{equation}  \label{aPn}
	p_n=  \left(1-e^{-\beta \hbar \omega} \right) e^{-n\beta \hbar \omega},
\end{equation}
for the eigenstate labelled by $n$ with eigenvalue $\hbar\omega\left(n+\frac{1}{2}  -|\alpha|^2 \right)$.  The associated density operator is the thermal state at inverse temperature $\beta = 1/(k_B T)$ 
\begin{equation}  \label{arhotherm}
	\hat\rho_{\beta}   = \sum_{n=0}^\infty p_n \, \ket{\tilde{n}_{\alpha}} \bra{\tilde{n}_{\alpha}}.
\end{equation}
The work distribution for unitary evolution starting from this statistical mixture is given by Eq.~\eqref{eq:p(w)} using the weights in  \eqref{aPn}.
Using Eqs.~(\ref{awq2}), (\ref{ahox0}) and (\ref{aPn}) we can now compute work averages. For example, the average work is
\begin{equation}   \label{appavWn}
\braket{W}_{\{n\}} = \frac{(A\tau)^2}{2m},
\end{equation}
which is the difference in the energy expectation values at $t=\tau$ and $t=0$, as expected. The average exponentiated work is given by
\begin{align} 
	\braket{e^{-\beta W}}_{\{n\}} = & \exp\left\{-\frac{A^2\tau^2\beta}{2m} \right. \nonumber\\ 
	& \times \left. \left[1 - \frac{\hbar\omega\beta}{2} \cos^2(\omega \tau)\coth\left(\frac{\hbar\omega\beta}{2} \right) \right] \right\}.  \label{aexpWho}
\end{align}

Before discussing this result in more detail in Section \ref{sec:comparison}, we will first analyse the case of an initial thermal state when it is prepared as a statistical mixture of coherent states.

\subsection{Initial thermal mixture of coherent states}

Above we constructed the initial thermal state as a mixture of energy eigenstates, Eq.~\eqref{arhotherm}, and obtained the average exponentiated work Eq.~\eqref{aexpWho}. However, the same initial density operator can be written as a mixture of coherent states $| \eta \rangle$, i.e.
 \begin{equation}   \label{arhotherm2}
	 \hat\rho_{\beta} = \int \d^2\eta \, p_{\alpha}(\eta) \, | \eta \rangle \langle \eta |,
\end{equation}
where $p_{\alpha}(\eta)$ is the $P$-representation~\cite{mandel,carmichael2009}. For the thermal state of the harmonic oscillator one finds~\cite{mandel,carmichael2009}
\begin{equation}    \label{aphi}
 	p_{\alpha}(\eta) = \frac{ \left(e^{\beta\hbar \omega} -1 \right) }{\pi}   \exp\left\{ -|\eta-\alpha|^2\left[e^{\beta\hbar \omega } -1 \right] \right\},
\end{equation}
where the Gaussian distribution has been shifted by $\alpha$ since the initial Hamiltonian $\hat{H}( \hat{x}, \hat{p}, 0)$ at time $t=0$, see Eq.~\eqref{aHho}, is displaced by $\hat D(\alpha)$ with energy eigenstates $\ket{\tilde{n}_{\alpha}}$. 
The different initial mixture in Eq.~(\ref{arhotherm2}) will give different quantum trajectories for the particle compared to the mixture in Eq.~(\ref{arhotherm}), even though both mixtures correspond to the same density operator. To obtain the work distribution and work averages for the coherent-state mixture we must find the trajectories for an initial coherent state.

For the oscillator initially in a coherent state $|\eta\rangle$ with complex amplitude $\eta$, with $\eta_R$ ($\eta_I$)  the real (imaginary) part of $\eta$, the state at $t>0$ is found from Eq.~(\ref{aT2}) to be $\exp\left(i \frac{At \eta_I}{\sqrt{2\hbar m\omega}}  \right)|\upsilon \rangle$, where $|\upsilon \rangle$ is a coherent state with time-dependent complex amplitude $\upsilon= \eta+\frac{At}{\sqrt{2\hbar m\omega}}$. The corresponding wave-function has the phase 
\begin{align}
	S(x,t)= & - \frac{1}{2} \hbar\omega t  + \frac{\hbar A t \eta_I}{\sqrt{2\hbar m\omega}}  \nonumber\\
			& - x\left[ \left(At+ \eta_R \sqrt{2\hbar m \omega} \right) \sin(\omega t) - \eta_I \sqrt{2\hbar m \omega} \cos(\omega t)\right]  \nonumber  \\ 
			& - \hbar \left(   \eta_R + \frac{At}{\sqrt{2\hbar m\omega}}  \right) \eta_I   \cos(2\omega t)  \nonumber\\
			& + \frac{\hbar}{2} \left[   \left(   \eta_R + \frac{At}{\sqrt{2\hbar m\omega}}  \right) ^2  -  [\eta_I]^2  \right] \sin(2\omega t).
\end{align}
This phase gives the particle trajectories
\begin{align}
	x_\eta(t)= 	& x_0- \eta_R \sqrt{\frac{2\hbar}{m\omega} }+ \left( \frac{At}{m\omega} + \eta_R \sqrt{\frac{2\hbar}{m\omega} } \right)  \cos(\omega t) \nonumber\\
				& + \eta_I \sqrt{\frac{2\hbar}{m\omega} }  \sin(\omega t).
\end{align}
For this initial wave function the work done on the oscillator between $t=0$ and $t=\tau$ is given by
\begin{align}  
W[x_\eta(t)]= & A \tau       \left[ \frac{A \tau}{2m}   +   \sqrt{\frac{2\hbar\omega}{m}}  \eta_R \right] + \left\{ A [ \omega \tau   \cos(\omega \tau) +\sin(\omega \tau) ] \right. \nonumber \\
			  & \left. + \sqrt{2\hbar m\omega^3}  \left[  \eta_R ( \cos(\omega \tau)-1) +   \eta_I \sin(\omega \tau)  \right]\right\} \nonumber\\
			  & \times\left[  x_0 -  \eta_R \sqrt{\frac{2\hbar}{m\omega}} \right]  .    \label{aWeta}
\end{align}
The probability distribution for the initial position $x_0$ is given by $|\psi_\eta(x_0)|^2$, where $\psi_\eta(x_0)$ is the wave function for the coherent state $|\eta\rangle$ at $t=0$:
\begin{equation} 
	|\psi_\eta(x_0)|^2 = \sqrt{\frac{m\omega}{\pi\hbar}} \exp\left\{- \frac{m\omega}{\hbar} 
	\left[ x_0 - \eta_R  \sqrt{\frac{2\hbar}{ m\omega}}  \right]^2 \right\}.  \label{apsisqdeta}
\end{equation}
The work distribution for the initial coherent state $|\eta \rangle$ is now determined by Eq.~\eqref{PtauW} with $\psi_\eta(x_0)$ as the initial state.

The work distribution arising for an initial thermal state $\hat\rho_{\beta} $ when it was prepared as a mixture of coherent states, \eqref{arhotherm2}, is given by 
\begin{equation}
	P(W)_{\{ \eta\}} = \int d^2\eta \,  p_{\alpha}(\eta)  P(W; \psi_\eta)
\end{equation}
with $ p_{\alpha}(\eta)$ given by Eq.~(\ref{aphi}). Work averages for this mixture can now be evaluated with Eqs.~(\ref{aphi}), (\ref{aWeta}) and (\ref{apsisqdeta}).

\subsection{Comparison of work averages for two thermal state mixtures.}\label{sec:comparison}

The average work for the initial coherent-state mixture is $\braket{ W}_{\{ \eta\}}=(A\tau)^2/(2m)$, which agrees with  Eq.~(\ref{appavWn}), as expected. However, when calculating the average exponentiated work $\braket{e^{-\beta W}}_{\{ \eta\}}$ one finds that it differs from the one obtained for an initial energy-state mixture in \eqref{aexpWho}. 
In fact, the quantity $\braket{e^{-\beta W}}_{\{ \eta\}}$ can diverge for certain values of the parameters, in contrast to the result for $\braket{e^{-\beta W}}_{\{ n\}}$. At high temperature $\braket{e^{-\beta W}}_{\{ \eta\}}$ always converges and to leading order in $\beta$ it is given by
\begin{equation} 	\label{appexpWetaex}
	\braket{e^{-\beta W}}_{\{ \eta\}}  = 1 +  \beta \hbar\omega \sin^2\left(\frac{\omega\tau}{2}\right)  
	+O(\beta^2),
\end{equation} 
whereas the high temperature limit of Eq.~\eqref{aexpWho} is
\begin{equation}  \label{appexpWnex}
	\braket{e^{-\beta W}}_{\{n\}}  = 1-\frac{\beta(A\tau)^2}{2m}\sin^2(\omega\tau)+O(\beta^2).
\end{equation} 
Note that to first order in $\beta$, $\braket{e^{-\beta W}}_{\{ \eta\}}$ is independent of the driving ($A$) whereas $\braket{e^{-\beta W}}_{\{ n\}}$ is not. The dependence on the driving in the high-temperature expansion of $\braket{e^{-\beta W}}_{\{ \eta\}}$ only appears at third order in $\beta$. The fact that moments of the work for a coherent-state mixture can be non-zero even without driving follows from the properties of the coherent state in the Hamilton-Jacobi formulation~\cite{holland1995quantum}. Although the average over all trajectories of the energy change of the particle in a pure (undriven) coherent state is zero, the energy change along individual trajectories is not zero in general, due to the quantum potential~\cite{holland1995quantum}. This means that higher moments of the energy change do not have to vanish.

For the time-dependent Hamiltonian (\ref{Hho}) the partition function associated with the thermal states is the same at any time and thus the classical Jarzynski equality is $\braket{e^{-\beta W}} = 1$. This is correctly obtained for both statistical mixtures, Eq.~\eqref{appexpWnex} and Eq.~\eqref{appexpWetaex}, in the high-temperature limit of $\beta\to 0$, which is the relevant classical limit here. 
We note that there is no {\it a priori} reason why the mathematical limit $\hbar\rightarrow 0$ should give the classical physical result here. Indeed, assuming  no contributions arising from higher orders of $\beta$, for the coherent-state mixture, \eqref{appexpWetaex}, the limit $\hbar \rightarrow 0$ happens to coincide with the classical value of 1. But for the energy-state mixture, \eqref{appexpWnex}, the $\hbar \rightarrow 0$ limit does not give 1, unless one substitutes the coupling constant $A$ as $A^2 = 2m \, \alpha_I^2 \, (\hbar \omega) \omega^2$ where $\alpha_I$ is the imaginary part of the displacement parameter $\alpha$.

The expressions show that in general there is no relation between the average exponentiated work $\braket{e^{-\beta W}}$ and equilibrium quantities, such as $\Delta F$, in the quantum case. 
This arises because each wave function already contributes an ensemble of trajectories to the work distribution in Eq.~\eqref{PtauW} and such ensembles are not related to Boltzmann distributions in any direct manner. 
Only the mixing over different wave functions can introduce thermal weights, and the resulting phase space distribution $ \rho(\bfx,\bfp,t) =  \sum_j p_j |\psi\uj(\bfx,t)|^2 \delta(\bfp - \nabla S^{\psi\uj}(\bfx,t)) $ is thus in general not of Boltzmann form, even if the mixture of $ \{\psi\uj\} $ corresponds to a thermal state $ \hat\rho_\beta $.

\section{Conclusion and Discussion.}  \label{sec:conclusion}

Quantum mechanics is routinely taught in a way that stresses subjectivity and indeterminism, while abolishing trajectories in phase space. As Bell pointed out, this is a ``deliberate theoretical choice'' that is ``not forced on us by experimental facts''~\cite{Bell1982}. Using quantum Hamilton-Jacobi theory we have here generalised the classical definition of work to the quantum case in a straightforward manner. We have provided a positive definite probability distribution characterising the work fluctuations in complete generality, including for coherent quantum systems. The distribution gives the average work as the change of energy expectation values for closed systems while recovering the classical result in the presence of classical dynamics.
While this may appear in contradiction with a recently proven ``no-go'' theorem~\cite{Perarnau17}, note that the work distribution (\ref{eq:p(w)}) is not linear in the density operator of the system. Thus, it is not of the form considered in Ref.~\cite{Perarnau17} and the theorem does not exclude this case.

In contrast to proposals based on trajectories in Hilbert space, such as the quantum jump approach~\cite{Elouard17,Hekking:2013aa,Pekola2013,Suomela:2014aa,Horowitz2012}, consistent histories framework~\cite{Miller2016} or continuous measurements~\cite{Alonso2016,Venkatesh2015}, the Hamilton-Jacobi approach is based on trajectories in phase space where the system has a well defined energy at all times. These trajectories have been reconstructed using weak measurements~\cite{Kocsis2011, Mahler2016, Xiao2017}, from which the distribution (\ref{eq:p(w)}) can be inferred. 

We here focussed on initial statistical mixtures undergoing closed dynamics - but the work distribution defined in \eqref{eq:p(w)} extends naturally to correlated and open systems. Instead of using the wave function of the system alone, the open case requires the use of the wave function $\psi_{\rm tot}$ of the system and any correlated or interacting degrees of freedom. The dynamics of the full system would have to be solved and the power integrated along the phase-space trajectory of the subsystem of interest.

Measurements are a particular example of open dynamics, where the full system consists of the measured system plus the measuring device. The Hamilton-Jacobi approach thus has the ability to explicitly quantify the work done during a measurement. Furthermore, it can be applied to initial states with coherences in the energy basis, whether or not measurements are performed. The removal of initial coherences through certain open processes has been found to allow work extraction~ \cite{Kammerlander2016}. This extracted work could be compared to the average work obtained from the work distribution (\ref{eq:p(w)}). More generally, it will be interesting to investigate how much work is done on the system when energy or other observables are measured.

We note that, as a general feature of the Hamilton-Jacobi formulation~\cite{holland1995quantum,DurrQuantum12}, the predictions for measurement outcomes derived from the Hamiltonian-Jacobi approach will agree with those of standard quantum mechanics. This includes standard tools of quantum mechanics, such as positive valued operator measurements (POVMs). Our work distribution is not associated with a POVM. This explains why here statistical mixtures rather than density operators determine the work distribution, in contrast to  previous work definitions. The broader perspective taken here offers new insights into current challenges in quantum thermodynamics by indicating that the statistical mixture may play a bigger role than usually anticipated. It will be interesting to explore implications of this feature on interpretational  aspects of quantum theory \cite{Cabello16,Carina17}.



As demonstrated in the harmonic oscillator example, the average exponentiated work in the quantum case is not  in general related to equilibrium properties such as the free energy change of the system. This difference from classical statistical physics occurs because the quantum thermal phase-space distribution is not of Boltzmann form and there is an explicit dependence on the statistical mixture.  Future investigations can explore how fluctuation relations can be generalized within the quantum Hamilton-Jacobi theory.

In the Hamilton-Jacobi approach the classical limit is transparent as it occurs whenever the quantum potential can be neglected in (\ref{eq:EBohm2}). The classical equations of motion and  corresponding trajectories are then recovered. In some cases the high temperature limit may also give classical results; and we showed that for the specific harmonic oscillator example the classical Jarzynski equality is recovered in this limit. An open problem is to find the general conditions under which high temperature gives classical statistical results. 

Many questions remain open in the field of quantum thermodynamics. The approach introduced here shows that they are firmly located in the realm of the ``speakable''~\cite{Bell1982}. 


\medskip

\textit{Acknowledgements.} We would like to thank M. Campisi, I. Starshinov, H. Miller, P. Muratore-Ginanneschi and B. Donvil for inspiring discussions and M. Campisi, J. Garrahan, C. Jarzynski, P. Solinas and P. Talkner for critical reading of the manuscript. This work has been in part supported by the Academy of Finland through its CoE grants 284621 and 287750. R.S. acknowledges support from the Magnus Ehrnrooth Foundation. J.A. acknowledges support from EPSRC (grant EP/M009165/1) and the Royal Society. This research was supported by the COST network MP1209 ``Thermodynamics in the quantum regime".

\bibliography{PRANov2017}

\end{document}